\documentclass[10pt]{article} 
\usepackage[english]{babel}
\usepackage[utf8]{inputenc}
\usepackage{fancyhdr}
\usepackage{mathtools}
\usepackage{comment}
\usepackage[symbol*]{footmisc}
\usepackage{enumerate}
\usepackage{geometry} 
\usepackage{amsmath}
\usepackage{amsmath}
\usepackage{amssymb}
\usepackage{upgreek}

\usepackage[utf8]{inputenc}
\usepackage{textcomp}
\usepackage{times}
\usepackage{lipsum}

\usepackage{graphicx} 
\usepackage{caption}
\usepackage[font=footnotesize] {caption}
\usepackage[font=footnotesize] {subcaption}

\usepackage{subcaption}
\usepackage{float} 
\restylefloat{figure}
\usepackage{wrapfig} 
\DeclareGraphicsExtensions{.pdf,.jpeg,.png}
\usepackage{lipsum} 
\usepackage{sectsty}

\usepackage[normalem]{ulem}
\usepackage{color}
\usepackage{comment}

\sectionfont{\large}
\subsectionfont{\normalsize}
\begin{document}\thispagestyle{empty}




\begin{center}
{\Large \bfseries A coupled mixed-mode cohesive zone model: An extension to\\ three-dimensional contact problems
\\[0.5cm]} 
\normalsize  Mohsen Khajeh Salehani\footnote[1]{Corresponding author.\\E-mail address: m.khajehsalehani@tudelft.nl.} and Nilgoon Irani\\[0.2cm]
\textit{Department of Materials Science and Engineering, Delft University of Technology, \\ 2628CD Delft, The Netherlands}
\normalsize \\ [0.2cm] (Dated: \today)
\end{center}
In this study a phenomenological three-dimensional coupled (3DC) mixed-mode cohesive zone model (CZM) is proposed. This is done by extending an improved version of the well established exponential CZM of Xu and Needleman (XN) to 3D contact problems. Coupled traction-separation relationships are individually presented for normal and transverse directions. The proposed model preserves all the essential features of the XN model and yet correctly describes mixed-mode separation and in particular mixed-mode closure conditions. Moreover, it provides the possibility to explicitly account for all three components of the gap function, i.e. separations in different directions. The 3DC model has some independent parameters, i.e. interface properties, similar to the XN model. All the cohesive zone parameters can be determined using mode-I and mode-II experiments. 
\section{Introduction}

Cohesive zone models (CZMs) were pioneered by Dugdale \cite{174} and Barenblatt \cite{173} and have been extensively employed to study the delamination process. In CZMs traction-separation laws are employed to describe the interface interactions as well as any associated dissipation. These models can be coupled or uncoupled. In an uncoupled model, the surface tractions are only dependent on their corresponding gap value \cite{162}. However, Savkoor and Briggs \cite{124} experimentally showed that in adhesive contact problems there is an interaction between different components of surface forces. As a result, in the last decades, increasing attention has been devoted to the use of coupled mixed-mode CZMs. The advantages of these methods over other approaches are their computational efficiency and versatility for numerical implementation \cite{83}.

In the relevant literature, there is a large variety of cohesive zone models. Most of them can be categorized into the following groups: polynomial \cite{164}, piece-wise linear \cite{165}, exponential \cite{88}, and rigid-linear cohesive zone models \cite{166} (for more details on each model see \cite{81}). Among them, the exponential type is one of the most popular CZMs due to the following advantages: First of all, a phenomenological description of contact is automatically achieved in normal compression. Secondly, the tractions and their derivatives are continuous, which is favourable from an implementation and computational point of view. Thirdly, the exponential models originate from the universal relationship between binding energies and atomic separation of bimetallic interfaces \cite{167}.

In addition to mixed-mode separation, CZMs have been also used to model mixed-mode closure. The term closure refers to the phenomena whereby contacting surfaces push into one another under a compressive load, e.g. in an indentation contact problem \cite{163}. However, physically realistic closure behaviour is not trivially achieved by most CZM formulations \cite{87}. 

In the present study, we focus on the exponential CZMs because of the advantages mentioned above. First, some existing models are reviewed. Then, it is shown that application of these models are limited to separation and/or two-dimensional cases. Finally, a coupled mixed-mode CZM in a three-dimensional setting is introduced which is applicable to the cases of separation and in particular closure.
\section{Some existing models and their range of applicability}
Among the widening class of exponential CZMs, the 2D model of Xu-Needleman (XN) is one of the most frequently used \cite{88}. In this model, at the cohesive surface, interfacial tractions in normal ($T_{1}$) and transverse directions ($T_{2}$) are defined as follows:
\begin{equation}
\setlength{\jot}{10pt}
\begin{split}
T_{1} &=  \dfrac{\phi_1}{\delta_1} ~ \text{exp} (- \dfrac{\Delta_1}{\delta_1})~\left\lbrace   \dfrac{\Delta_1}{\delta_1} ~  \text{exp} ( - \dfrac{\Delta_2^{2}}{\delta_2^{2}} ) + \dfrac{1-q}{r-1} \left[ 1 - \text{exp} ( - \dfrac{\Delta_2^{2}}{\delta_2^{2}} ) \right] \left[ r -  \dfrac{\Delta_1}{\delta_1} ~ \right]  \right\rbrace, \\
T_{2} &=2~ \dfrac{\phi_2}{\delta_2} ~ \dfrac{\Delta_2}{\delta_2} ~ \left[ q +  \dfrac{r-q}{r-1} ~ \dfrac{\Delta_1}{\delta_1} \right]  \text{exp} ( - \dfrac{\Delta_1}{\delta_1} ) ~\text{exp} ( - \dfrac{\Delta_2^{2}}{\delta_2^{2}}),
\end{split}
\label{XNn}
\end{equation}
where $\phi_i$, $\delta_i$ and $\Delta_i$ are work of separation, characteristic length and gap value in direction $i=\{1, 2\}$, respectively. Coupling in this model is controlled through the parameters $q = \phi_2 / \phi_1$ and $r = \Delta^{*}_1 / \delta_1$, where $\Delta^{*}_1$ is the value of $\Delta_1$ after complete separation in transverse direction takes place under the condition of $T_{1} = 0$.

Since Xu and Needleman introduced their exponential CZM in 1993, their model has been altered and extended by several authors. In 2006, a comprehensive study on the coupling parameters, $r$ and $q$ was performed by van den Bosch \textit{et al.} \cite{81}. They demonstrated that only for $r = q$ the required transverse traction increases with increasing compression in the normal direction. Moreover, they showed that the XN model yields to unrealistic results for $r \neq 1$, since even after complete separation in transverse direction, $T_{1}$ does not become zero. Hence, they concluded that only for $r = q = 1$ a physical coupling behaviour is obtained. Nevertheless, setting $q$ equal to unity causes other issues. First, choosing $q = 1$ implies that $\phi_1 = \phi_2$. This assumption is often made in literature however, multiple experiments show that work of separation in different directions are not equal \cite{93,94}. For solving this issue, an adjusted model based on the XN model was proposed by van den Bosch \textit{et al.} \cite{81}, the BSG model:
\begin{equation}
\setlength{\jot}{10pt}
\begin{split}
T_{1} &=  \dfrac{\phi_1}{\delta_1} ~\dfrac{\Delta_1}{\delta_1}~ \text{exp} (- \dfrac{\Delta_1}{\delta_1})~ \text{exp} ( - \dfrac{\Delta_2^{2}}{\delta_2^{2}} ) , \\
T_{2} &=2~ \dfrac{\phi_2}{\delta_2} ~ \dfrac{\Delta_2}{\delta_2} ~ \left[ 1 +  \dfrac{\Delta_1}{\delta_1} \right]  \text{exp} ( - \dfrac{\Delta_1}{\delta_1} ) ~\text{exp} ( - \dfrac{\Delta_2^{2}}{\delta_2^{2}}).
\end{split}
\label{BSG}
\end{equation}
The BSG model works perfectly for problems with mixed-mode separation. However, setting $q$ equal to unity also raises an issue in problems involving mixed-mode closure: the work done in transverse direction reduces with increasing compressive load in normal direction. Consequently, for large values of closure $\Delta_{1}/\delta_{1}<-1$, a negative work for transverse direction is computed.

In order to correct the non-physical behaviour of the BSG model in mixed-mode closure problems, McGarry \textit{et al.} \cite{87} proposed a modified form of the BSG traction–separation relationship, the so-called NP1 model. This is done by simply taking out the term $[ 1+\Delta_{1}/\delta_{1} ]$ from the definition of $T_{2}$ in Eq. (\ref{BSG}), without changing the relation for $T_{1}$.
\begin{equation}
\setlength{\jot}{10pt}
\begin{split}
T_{1} &=  \dfrac{\phi_1}{\delta_1} ~\dfrac{\Delta_1}{\delta_1}~ \text{exp} (- \dfrac{\Delta_1}{\delta_1})~ \text{exp} ( - \dfrac{\Delta_2^{2}}{\delta_2^{2}} ) , \\
T_{2} &=2~ \dfrac{\phi_2}{\delta_2} ~ \dfrac{\Delta_2}{\delta_2} ~  \text{exp} ( - \dfrac{\Delta_1}{\delta_1} ) ~\text{exp} ( - \dfrac{\Delta_2^{2}}{\delta_2^{2}}).
\end{split}
\label{NP1}
\end{equation}
This new formulation is able to obtain correct coupling in both separation and closure problems.

Here, from the large body of literature few CZMs were addressed in a two-dimensional setting. Contrary to 2D cases, only a relatively small number of studies have been devoted to 3D cohesive surfaces \cite{168,169,170}. In one class of these models an effective gap value $\Delta_{\text{eff}}$ is employed:
\begin{equation}
\Delta_{\text{eff}} = \sqrt{\Delta_{1}^{2}+\beta^{2}(\Delta_{2}^{2}+\Delta_{3}^{2})},
\label{delta_eff}
\end{equation}
where $\beta$ is a scalar and assigns a different weight to opening displacements for transverse and normal directions. Subsequently, the calculated $\Delta_{\text{eff}}$ is employed to obtain the traction value. Its merits include simplicity and a straightforward formulation in 3D. However, whether these models can be applied in mixed-mode problems remains an issue \cite{171}. Besides, to the best of our knowledge, these models are merely applied into crack opening problems, i.e. single-mode separation cases. 

In another class of 3D CZMs, e.g. \cite{171,172}, all three components of the gap function are explicitly incorporated into the calculation of surface tractions. These models follow the framework of Xu and Needleman. Hence, they are very similar to Eq. (\ref{XNn}) or (\ref{BSG}) and work perfectly in separation problems. However, they show unrealistic behaviour in indentation-induced problems, as discussed earlier in the BSG model. 

\section{Extension to 3D contact problems: 3DC model}
In an effort to overcome the above-mentioned problems of the existing models, we propose a phenomenological three-dimensional coupled (3DC) mixed-mode CZM. Since the NP1 model \cite{87} of Eq. (\ref{NP1}) works perfectly in both problems of separation and closure, the proposed 3DC model will be obtained by extending NP1 to 3D contact problems.

In the 3DC model, the traction-separation relationship $T_{i}$ in normal ($\text{x}_{1}$) and transverse directions ($\text{x}_{2},\text{x}_{3}$) have the following expressions:
\begin{equation}
T_{i} = \dfrac{\phi_{i}}{\delta_{i}}~\dfrac{\Delta_{i}}{\delta_{i}}~\text{exp}\left[ - \sum_{j=1}^{d} (\dfrac{\Delta_{j}}{\delta_{j}})^{\alpha} \right]; \hspace{8mm} i=[1,d ],\hspace{5mm} \alpha = 
\left\{ \begin{array}{lcl}
1 \hspace{8mm} j=1 \\
2 \hspace{8mm} j \neq 1
\end{array}
\right.
,
\label{3DC}
\end{equation}
where $d=3$ is the dimension of the problem. With transverse isotropic assumptions, work of separation and characteristic length for both transverse directions are the same:
\begin{equation}
\setlength{\jot}{10pt}
\begin{split}
\phi_2 &= \phi_3 \coloneqq \phi_\text{t} , \\
\delta_2 &= \delta_3 \coloneqq \delta_\text{t}.
\end{split}
\label{istp}
\end{equation}
Note that by limiting the dimension and putting $d=2$ for a 2D case, the proposed model in Eq. \ref{3DC} almost coincides with the formulation of Eq. (\ref{NP1}). Hence, it provides a physically realistic representation of the cases of mixed-mode separation and in particular mixed-mode closure. Moreover, it preserves all the essential features of the original XN model. Similar to the XN model, the 3DC model is built upon different independent parameters, i.e. $\phi_{i}$ and $\delta_{i}$, which have physical meaning and can be determined via mode-I (opening mode) and mode-II (shearing mode) experiments.

\begin{figure}[!htp]
\begin{subfigure}{.5\textwidth}
  \centering
  \includegraphics[width=0.85\linewidth]{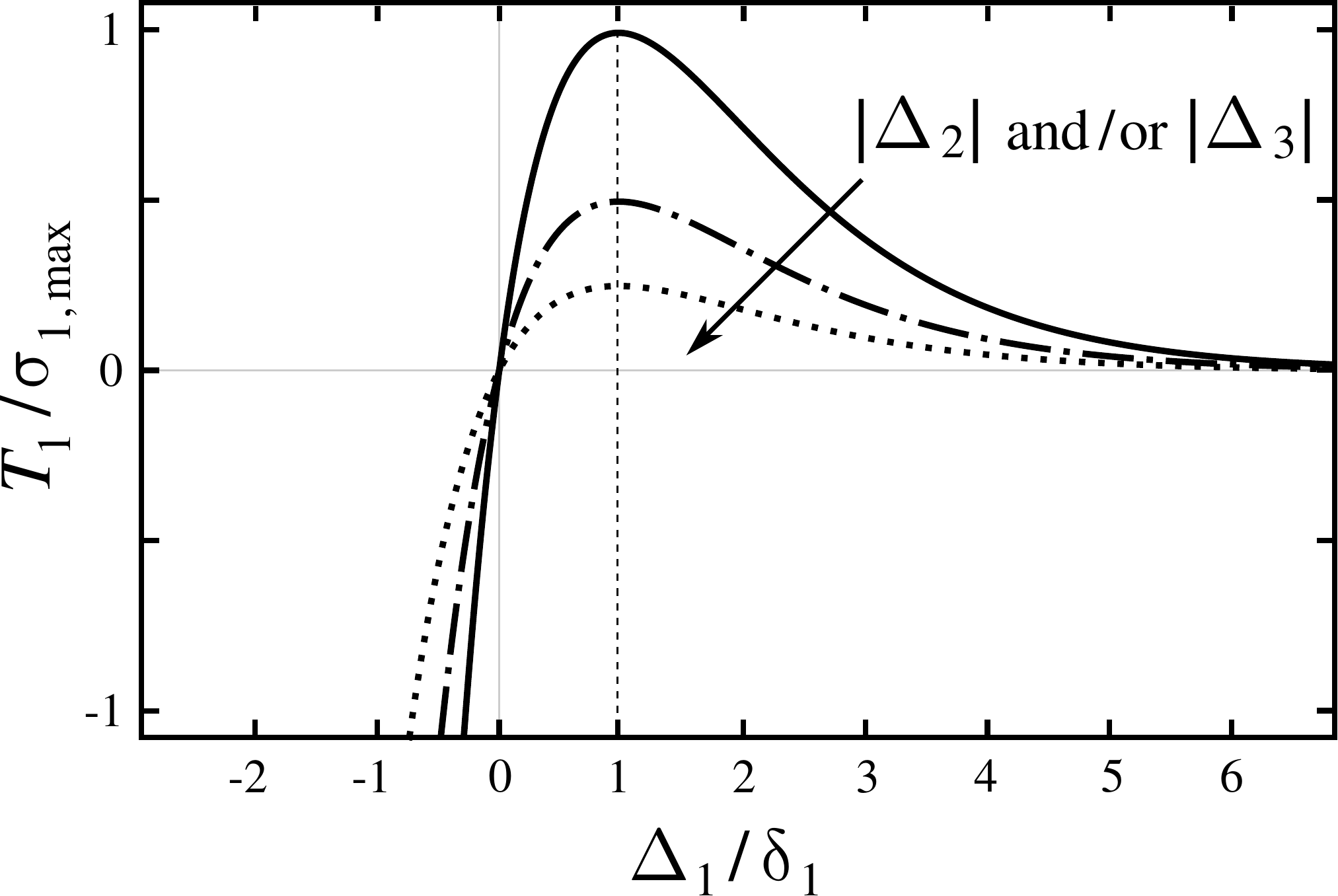}
  \caption{\centering normal direction: $\text{x}_{1}$}
  \label{norm}
\end{subfigure}%
\begin{subfigure}{.5\textwidth}
  \centering
  \includegraphics[width=0.85\linewidth]{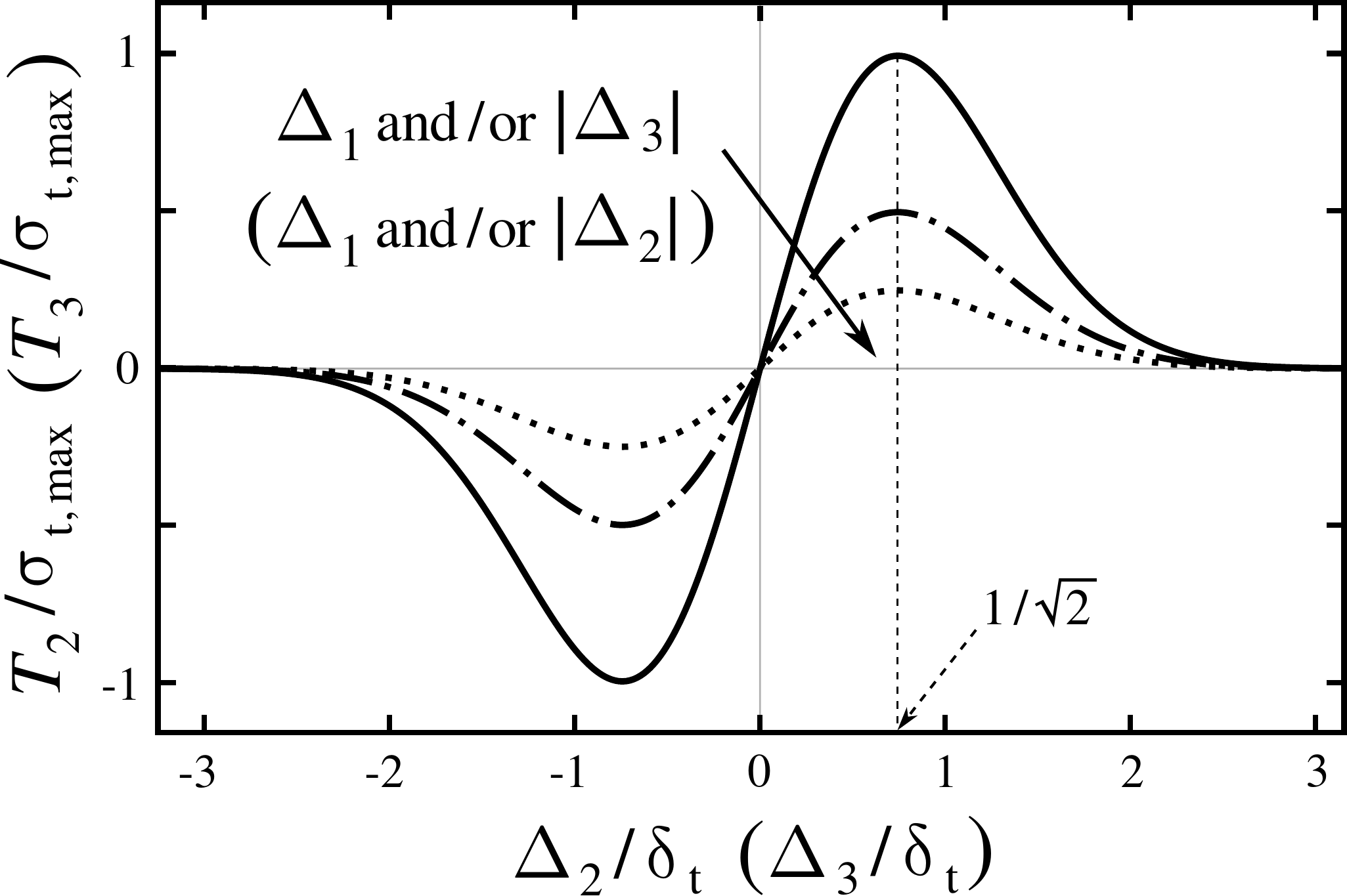}
  \caption{\centering transverse direction: $\text{x}_{2}$ ($\text{x}_{3}$)}
  \label{lat}
\end{subfigure}
\caption{Graphical representation of normalized coupled tractions versus normalized gap values as given by Eq. (\ref{3DC}).}
\label{3DCCZM}
\end{figure}

The maximum value of the traction $T_{i,\text{max}}$ and its corresponding gap value is given as:
\begin{equation}
T_{i,\text{max}} \vert_{\Delta_{i} = \delta_{i} / \eta} = \sigma_{i,\text{max}}~\text{exp}\left[ - \sum_{j=1, ~ j\neq i}^{d} (\dfrac{\Delta_{j}}{\delta_{j}})^{\alpha} \right]; \hspace{8mm} \eta = 
\left\{ \begin{array}{lcl}
\hspace{1.5mm}1 \hspace{9mm} i=1 \\
\sqrt{2} \hspace{7mm} i \neq 1
\end{array}
\right.
,
\label{3DC_max}
\end{equation}
where $\sigma_{i,\text{max}}$ is the strength of the interface in $i-$direction without considering separation in other directions, i.e. uncoupled strength of each individual mode. The following relation holds for $\sigma_{i,\text{max}}$:
\begin{equation}
\sigma_{i,\text{max}} = \dfrac{1}{\lambda} ~ \dfrac{\phi_{i}}{\delta_{i}} ; \hspace{8mm} \lambda = 
\left\{ \begin{array}{lcl}
\hspace{2.5mm}e \hspace{9.9mm} i=1 \\
\sqrt{2e} \hspace{7.5mm} i \neq 1
\end{array}
\right.
,
\label{3DC_max_uncpl}
\end{equation}
where $e = \text{exp}(1)$ and $\sigma_{2,\text{max}} = \sigma_{3,\text{max}} \coloneqq \sigma_{\text{t},\text{max}}$. The coupled tractions versus corresponding gap values are graphically shown in Fig.~\ref{3DCCZM}.

For modelling a general anisotropic material, the proposed 3DC model must be extended to also accurately account for the mode-III (tearing mode) as well as mode-I and mode-II besides their mode-mixity, but this is beyond the scope of the current work. 
\section{Concluding remarks}
We have focused our attention on the coupled mixed-mode exponential cohesive zone models (CZMs). A phenomenological three-dimensional coupled (3DC) CZM is proposed which is applicable to separation and in particular closure mixed-mode problems. To this end, an improved version of the well established Xu and Needleman model has been extended to 3D contact problems. 

The proposed 3DC model provides the possibility to explicitly account for all three components of the gap function. This model, similar to the XN model, is built upon different independent parameters, i.e. interface properties. All the cohesive zone parameters can be determined using mode-I and mode-II experiments. The proposed model can be applied to study the delamination process, crack propagation, thin film peeling, and several other applications where mode-mixity comes into play.

Although this model provides a fundamental platform for modelling the interface in three-dimensional contact problems, further verification is yet needed through experimental tests and computer simulations.
{\small
\bibliographystyle{ieeetr}
\bibliography{bib}
}
\end{document}